# Failure of Numerical modeling of 3-D Position Reconstruction from 3-Axial Planar Spiral Coil Sensor Sensitivity due to Existence of Quadratic Terms

Edi Sanjaya[1], Mitra Djamal[2], and Sparisoma Viridi[3,*]

[1]Doctoral Program in Physics, Institut Teknologi Bandung, Bandung 40132
[2]Theoretical High Energy Physics and Instrumentation Research Division,
Institut Teknologi Bandung, Bandung 40132, Indonesia
[3]Nuclear Physics and Biophysics Research Division, Institut Teknologi Bandung, Bandung 40132, Indonesia
endong_s@yahoo.com, mitra@fi.itb.ac.id, dudung@fi.itb.ac.id
*Corresponding author

Abstract

A sensitivity profile of a planar spiral coil sensor (PSCS) is proposed and is used to generate the relation of 3-D position of object observed using three PSCSs, one in each $x$, $y$, and $z$ axis to the sensors response. A numerical procedure using self consistent field-like method to reconstruct the real position of observed object from sensor sensitivity is presented and the results are discussed. Unfortunately, the procedure fails to approach the desired results due to the existence of quadratic terms.

Abstrak

Profil sensitivitas sebuah sensor koil datar atau planar spiral coil sensor (PSCS) diusulkan dalam tulisan ini dan digunakan untuk menghasilkan relasi dalam 3-D antara posisi obyek dengan menggunakan tiga PSCS, satu dalam setiap sumbu $x$, $y$, dan $z$ terhadap repons sensor. Prosedur numerik dengan mengunakan metode mirip self consisten field untuk merekonstruksi posisi sebenarnya obyek yang teramati dari informasi sensitivitas sensor dilaporkan dan didiskusikan. Sayangnya, prosedur yang diusulkan ini gagal untuk mencapai hasil yang diinginkan dikarenakan adanya suku-suku kuadratik.

Kata kunci: sensor koil datar, profil sensitivitas, 3-posisi, rekonstruksi numerik

## I. INTRODUCTION

The use of induction coils as sensors are very abroad nowadays with various constructions [1]. Among these constructions there is planar spiral coil sensor (PSCS) which is not so complex in design but powerful in detecting position and vibration [2]. It has been reported recently that this type of sensor can detect magnetic field as part of planar inductor [3], can act as an artificial tactile mechanoreceptor [4], can detect hidden metallic object as used as an array [5], and can also sense the forming of a metal sheet [6]. Unfortunately, its sensitivity depends on position of observed object to the sensor. Model of sensor sensitivity profile is proposed in this report and the relation to the 3-D position of observed object is presented.

## II. SENSITIVITY PROFILE

A model of PSCS sensitivity profile is proposed in this work, which depends on observed object position to the sensor. Since the geometry of PSCS lays on a planar surface it must have a cylindrical symmetry. It means that it can not differ two position of object when both positions have the same radial distance. An illustration is given in Figure 1 for plane size $L \times L$.

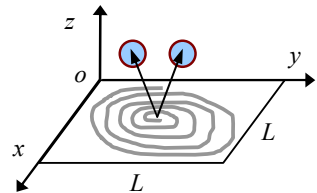

Figure 1. A PSCS will generate the same signal for both spherical objects since they have the same radial distance from center of the sensor.

A gaussian form is chosen sensor sensitivity in radial direction which can be formulated as

$$S(\rho) \propto \exp\left[-\frac{\rho^2}{2\sigma^2}\right]. \quad (1)$$

And for axial direction it also chosen in exponential form as

$$S(h) \propto \exp[-\gamma h]. \quad (2)$$

Then by introducing a value $S_0$ at position (0, 0, 0), Equation (1) dan (2) can written in an equation such as

$$S(\rho,h) = S_0 \exp\left[-\frac{\rho^2}{2\sigma^2}\right]\exp[-\gamma h]. \quad (3)$$

Since there will be three PSCS used in the model, it is necessary to define an equation that accommodate these three sensor, which are placed in $xy$, $yz$, and $zx$ plane, respectively. In order to do this, definition of a vector is needed

$$\vec{r} = x\hat{i} + y\hat{j} + z\hat{k}. \quad (4)$$

A sensor sensitivity that lays on a $xy$ or $ij$ plane is defined as

$$S_{ij}(x,y,z) = \\ S_0 \exp\left[-\frac{(\vec{r}\cdot\hat{i}-\frac{1}{2}L)^2 + (\vec{r}\cdot\hat{j}-\frac{1}{2}L)^2}{2\sigma^2}\right] \times \\ \exp\left[-\gamma(\hat{i}\times\hat{j})\cdot\vec{r}\right] \quad (5)$$

Then the sensitivity of sensor placed in $xy$, $yz$, and $zx$ plane for every position $(x,y,z)$ is given by $S_{ij}(x,y,z)$, $S_{jk}(x,y,z)$, and $S_{ki}(x,y,z)$, respectively. Figure 2. shows the illustration of three PSCSs in $xy$, $yz$, and $zx$ plane. Every PSCS will give different response, which depends on observed object position.

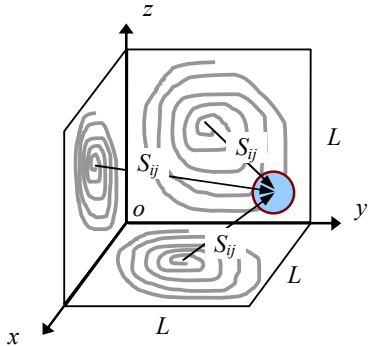

Figure 2. The spherical object position introduces three different PSCS responses, one response from each plane ($ij$, $jk$, and $ki$ plane).

## III. RECONSTRUCTION PROCEDURE

New variable $S_x$, $S_y$, and $S_z$ produced by the sensors, are introduced. A relation between ($S_x$, $S_y$, $S_z$) and ($x$, $y$, $z$) must be determined. The form of Equation (5) can be simplified using new variables, which are

$$S_x = S_0 \exp\left[-\frac{(y-\frac{1}{2}L)^2}{2\sigma^2}\right] \times \\ \exp\left[-\frac{(z-\frac{1}{2}L)^2}{2\sigma^2}\right]\exp(-\gamma x) \quad (6)$$

$$S_y = S_0 \exp\left[-\frac{(z-\frac{1}{2}L)^2}{2\sigma^2}\right] \times \\ \exp\left[-\frac{(x-\frac{1}{2}L)^2}{2\sigma^2}\right]\exp(-\gamma y) \quad (7)$$

$$S_z = S_0 \exp\left[-\frac{(x-\frac{1}{2}L)^2}{2\sigma^2}\right] \times \\ \exp\left[-\frac{(y-\frac{1}{2}L)^2}{2\sigma^2}\right]\exp(-\gamma z) \quad (8)$$

Following equation can be found directly from Equation (6)-(8)

$$\frac{(y-\frac{1}{2}L)^2}{2\sigma^2} + \frac{(z-\frac{1}{2}L)^2}{2\sigma^2} + \gamma x = \ln\left(\frac{S_0}{S_x}\right), \quad (9)$$

$$\frac{(z-\frac{1}{2}L)^2}{2\sigma^2} + \frac{(x-\frac{1}{2}L)^2}{2\sigma^2} + \gamma y = \ln\left(\frac{S_0}{S_y}\right), \quad (10)$$

$$\frac{(x-\frac{1}{2}L)^2}{2\sigma^2} + \frac{(y-\frac{1}{2}L)^2}{2\sigma^2} + \gamma x = \ln\left(\frac{S_0}{S_x}\right). \quad (11)$$

We can change the origin from (0, 0, 0) to $\left(\frac{1}{2}L, \frac{1}{2}L, \frac{1}{2}L\right)$ for simplicity, that leads Equation (9)-(11) to

$$x = \frac{1}{\gamma}\ln\left(\frac{S_0}{S_x}\right) - \frac{y^2}{2\gamma\sigma^2} - \frac{z^2}{2\gamma\sigma^2}, \quad (12)$$

$$y = \frac{1}{\gamma}\ln\left(\frac{S_0}{S_y}\right) - \frac{x^2}{2\gamma\sigma^2} - \frac{z^2}{2\gamma\sigma^2}, \quad (13)$$

$$z = \frac{1}{\gamma}\ln\left(\frac{S_0}{S_z}\right) - \frac{x^2}{2\gamma\sigma^2} - \frac{y^2}{2\gamma\sigma^2}. \quad (14)$$

The Equation (12)-(14), can be simplified in written them into a single equation, which is

$$\begin{pmatrix} x \\ y \\ z \end{pmatrix} = -\frac{1}{\gamma} \ln \left[ \frac{1}{S_0} \begin{pmatrix} S_x \\ S_y \\ S_z \end{pmatrix} \right] - \frac{1}{2\gamma\sigma^2} \begin{pmatrix} 0 & y^2 & z^2 \\ x^2 & 0 & z^2 \\ x^2 & y^2 & 0 \end{pmatrix}. \quad (15)$$

The Equation (15) can be solved numerically by giving guessing value of $(x_0, y_0, z_0)$ and calculate the result iteratively using a self consistent-like method.

The iterative algorithm to solve Equation (15) is as follow

1. start
2. guess initial condition $(x_0, y_0, z_0)$
3. set $x(t) = x_0, y(t) = y_0, z(t) = z_0$
4. calculate the value of $[x(t+1), y(t+1), z(t+1)]$ using Equation (15)
5. check whether the difference $\Delta x = |x(t+1) - x(t)|$, $\Delta y = |y(t+1) - y(t)|$, and $\Delta z = |z(t+1) - z(t)|$ smaller thatn a value, if yes go to step 8
6. set $x(t) = x(t+1), y(t) = y(t+1), z(t) = z(t+1)$
7. $t \to t + 1$ and repeat step 4
8. stop

## IV. RESULTS AND DISCUSSION

Set of parameters is chosen for the modeling, where $L$ = 2 cm, $\sigma$ = 0.5 cm, $S_0$ = 1 mV, and $\gamma$ = 50 m$^{-1}$. The influence of object position in axial direction $h$ is illustrated in Figure 3.

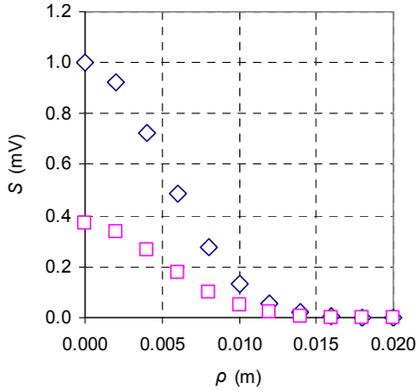

Figure 3. Profile of a PSCS sensitivity with $L$ = 2 cm, $\sigma$ = 0.5 cm, $S_0$ = 1 mV, and $\gamma$ = 50 m$^{-1}$ for: $h$ = 0 cm ($\diamond$) and $h$ = 2 cm ($\square$).

Nine points as sample of $(x, y, z)$ are calculated to see the influence of position to value of $S_{ij}$, $S_{jk}$, and $S_{ki}$ as in Table 1. It shows that position of observed object in a direction is given by sensitivity of a PSCS in a plane perpendicular to the direction, i.e. position in x or i direction is given by $S_{jk}$ since $j \times k = i$.

Table 1. Values of $S_{ij}(x,y,z)$, $S_{jk}(x,y,z)$, and $S_{ki}(x,y,z)$ for several value of $(x,y,z)$.

| No. | x/L | y/L | z/L | $S_{ij}$ (mV) | $S_{jk}$ (mV) | $S_{ki}$ (mV) | ΣS (mV) |
|---|---|---|---|---|---|---|---|
| 1 | 0.0 | 0.0 | 0.0 | 0.018 | 0.018 | 0.018 | 0.055 |
| 2 | 1.0 | 0.0 | 0.0 | 0.018 | 0.007 | 0.018 | 0.043 |
| 3 | 0.0 | 1.0 | 0.0 | 0.018 | 0.018 | 0.007 | 0.043 |
| 4 | 1.0 | 1.0 | 0.0 | 0.018 | 0.007 | 0.007 | 0.032 |
| 5 | 0.0 | 0.0 | 1.0 | 0.007 | 0.018 | 0.018 | 0.043 |
| 6 | 1.0 | 0.0 | 1.0 | 0.007 | 0.007 | 0.018 | 0.032 |
| 7 | 0.0 | 1.0 | 1.0 | 0.007 | 0.018 | 0.007 | 0.032 |
| 8 | 1.0 | 1.0 | 1.0 | 0.007 | 0.007 | 0.007 | 0.020 |
| 9 | 0.5 | 0.5 | 0.5 | 0.607 | 0.607 | 0.607 | 1.820 |

Then it can be understood why there are new variable introduced such as $S_x$, $S_y$, and $S_z$ in order to get value of $x$, $y$, $z$ from the sensors. In this work initial value of $\left(\frac{1}{2}L, \frac{1}{2}L, \frac{1}{2}L\right)$ is used or (0, 0, 0) for new origin as in Equation (15).

Direct application of (15) unfortunately can not get the value of $(x, y, z)$ from $S_x$, $S_y$, and $S_z$. The unexpected results are shown in Table 2 whose values for $(x, y, z)$ are different than previously chosen in Table 1.

Table 2. Values of $(x, y, z)$ .which are reconstructed from $S_{ij}(x,y,z)$, $S_{jk}(x,y,z)$, and $S_{ki}(x,y,z)$.

| No. | x/L | y/L | z/L | $S_{ij}$ (mV) | $S_{jk}$ (mV) | $S_{ki}$ (mV) | ΣS (mV) |
|---|---|---|---|---|---|---|---|
| 1 | -0.131 | -0.131 | -0.131 | 0.018 | 0.018 | 0.018 | 0.055 |
| 2 | -0.222 | -0.319 | -0.222 | 0.018 | 0.007 | 0.018 | 0.043 |
| 3 | -0.222 | -0.222 | -0.319 | 0.018 | 0.018 | 0.007 | 0.043 |
| 4 | -0.088 | -0.085 | -0.085 | 0.018 | 0.007 | 0.007 | 0.032 |
| 5 | -0.319 | -0.222 | -0.222 | 0.007 | 0.018 | 0.018 | 0.043 |
| 6 | -0.085 | -0.085 | -0.088 | 0.007 | 0.007 | 0.018 | 0.032 |
| 7 | -0.085 | -0.088 | -0.085 | 0.007 | 0.018 | 0.007 | 0.032 |
| 8 | -0.111 | -0.111 | -0.111 | 0.007 | 0.007 | 0.007 | 0.020 |
| 9 | +0.059 | +0.059 | +0.059 | 0.607 | 0.607 | 0.607 | 1.820 |

It means that a modification to (15) must be conducted in order to get the desired value of $(x, y, z)$ from $S_x$, $S_y$, and $S_z$.

Failure of the direct implementation of the Equation (15) could be sourced from the fact that $S_x$, $S_y$, and $S_z$ contains quadratic terms of $x$ and $y$, or $y$ and $z$, or $z$ and $y$, which could confuse the computation process as once reported in other case for $\cos\theta$ with $\theta \in [-\pi/2, \pi/2]$, where all values of $\cos\theta$ always positive and can not be traced back to sign of $\theta$ [7]. Because of this the mapping-back process is not unique.

## V. CONCLUSION

It has been shown that direct implementation of a numerical self consistent field-like method has failed to obtained $(x, y, z)$ from $S_x$, $S_y$, and $S_z$. Modification to or other approach to substitute the proposed method is needed. Source of failures can be address to the quadratic terms which makes mapping-back process not unique.

## ACKNOWLEDGMENT

S. V. would like to thank to Institut Teknologi Bandung Research Division Research Grant in 2010 for supporting computation part of this work partially.